\input{aipcheck}

\documentclass[
    ,final            
  ]
  {aipproc}

\layoutstyle{6x9}

\usepackage{amssymb}
\usepackage{graphicx}
\usepackage{natbib}
\usepackage{hyperref}
\usepackage{mathptmx}
\usepackage{sidecap}

\begin{document}

\title{Non thermal emission from clusters of galaxies: the importance of a joint LOFAR/Simbol-X view}

\classification{98.65.Cw,95.85.Bh,95.85.Nv}
\keywords      {cosmology: large-scale structure -- galaxies: clusters: general -- galaxies: intergalactic medium -- radiation mechanisms: non-thermal}

\author{C. Ferrari}{ address={UNSA, CNRS UMR 6202 Cassiop\'ee,
    Observatoire de la C\^ote d'Azur, Nice, France} }

\begin{abstract}
  Deep radio observations of galaxy clusters have revealed the
  existence of diffuse radio sources ("halos" and "relics") related to
  the presence of relativistic electrons and weak magnetic fields in
  the intracluster volume. I will outline our current knowledge about
  the presence and properties of this non-thermal cluster
  component. Despite the recent progress made in observational and
  theoretical studies of the non-thermal emission in galaxy clusters,
  a number of open questions about its origin and its effects on the
  thermo-dynamical evolution of galaxy clusters need to be answered. I
  will show the importance of combining galaxy cluster observations by
  new-generation instruments such as LOFAR and Simbol-X.  A deeper
  knowledge of the non-thermal cluster component, together with
  statistical studies of radio halos and relics, will allow to test
  the current cluster formation scenario and to better constrain the
  physics of large scale structure evolution.
\end{abstract}

\maketitle


\section{Introduction}

Galaxy clusters, the largest gravitationally bound virialized
structures in the Universe, have been discovered at optical
wavelengths as concentrations of hundreds to thousands of galaxies in
regions of a few Mpc. Most of the cluster baryonic matter is made up
by a hot ($T \approx 10^8$ K) and diffuse ($n \approx 10^{-3} {\rm
  cm^{-3}}$) intracluster medium (ICM), emitting in X-ray due to
thermal bremsstrahlung \cite{1988xrec.book.....S}. In the last 10
years, joint detailed optical and X-ray analyses
\citep[e.g.][]{2006A&A...446..417F} have shown that galaxy clusters
form and evolve through merging of less massive systems, in agreement
with the expectation of the hierarchical scenario of structure
formation emerging from the concordant cosmological model
($\Lambda$CDM). It has also been proved that more than 80\% of the
cluster mass is in the form of dark matter, only detectable through
its gravitational field. On the opposite, the last radiative component
of clusters is totally negligible in terms of mass. It is made up by
intracluster relativistic electrons and magnetic fields, whose origin
and role within cluster physics are still matter of debate (see
\cite{2008LNP...740..143F,2008SSRv..134...93F} and refs. therein).

In the following, I will give an overview of our current knowledge of
the non-thermal component of galaxy clusters. The perspectives that
will be opened in this field by the radiotelescope LOFAR and the X-ray
satellite Simbol-X will also be discussed. The $\Lambda$CDM model with
H$_0$=70 km ${\rm s}^{-1} {\rm Mpc}^{-1}$, $\Omega_m=0.3$ and
$\Omega_{\Lambda}=0.7$ has been adopted.

\section{The intracluster non-thermal component} 

\subsection{Observational evidence}

\begin{figure}
   \centering
   \includegraphics[width=2.2in,angle=-90]{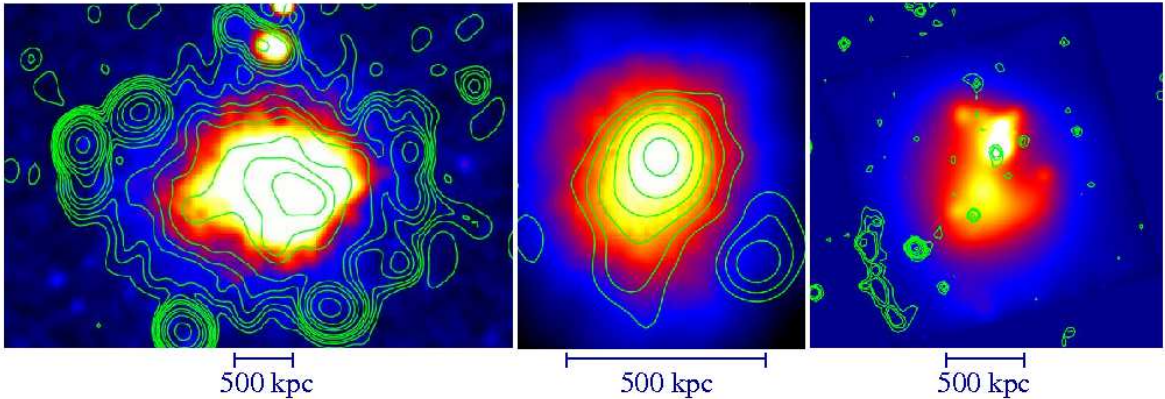} 
   \caption{Examples of the three classes of diffuse radio sources in
     galaxy clusters: halos (left), mini-halos (centre) and relics
     (right). Radio contours are overlaid on the X-ray images of the
     three galaxy clusters A2163, RX J1347-1145 and A521. Radio data
     are from the VLA
     \cite{2001A&A...373..106F,2007A&A...470L..25G,2006A&A...446..417F},
     X/ray data from XMM-Newton \cite{bourdin2163,2007A&A...472..383G}
     or Chandra observations \cite{2006A&A...446..417F}.}
   \label{fig:fig1}
\end{figure}

\noindent \textbf{Radio Observations -} Diffuse radio emission from
galaxy clusters was discovered in Coma by \citet{1959Natur.183.1663L}
and then confirmed by \citet{1970MNRAS.151....1W}, who suggested that
the extended synchrotron radio emission was due to the presence of
relativistic electrons (Lorentz factor $\gamma>>$1000) and magnetic
fields ($\mu$G) in the intracluster volume. Radio observations thus
provided the first direct evidence of the presence of a non-thermal
intracluster component. Statistical studies of the intracluster radio
emission could start only in the 90's
\citep[e.g.][]{1999NewA....4..141G} with the advent of continuum radio
surveys such as the NVSS \cite{1998AJ....115.1693C}. The surface
brightness of diffuse radio sources in clusters is very low
($\sim\mu$Jy$/{\rm arcsec}^2$ at 1.4 GHz). The detection of extended
cluster radio emission is thus a challenge and, at present, only about
60 clusters are known to host diffuse radio sources. These are all
characterised by steep synchrotron spectra, but can differ in their
morphology and position in the host clusters (see
Fig. \ref{fig:fig1}). The following working definition is usually
adopted a) {\it radio halos} are extended ($\gtrsim$1 Mpc) sources
that have been detected at the centre of merging clusters; their
morphology is similar to the X-ray morphology of the cluster; b) {\it
  radio mini-halos} are smaller sources ($\lesssim$500 kpc) located at
the centre of cooling flow clusters; they surround a powerful radio
galaxy; c) {\it radio relics} have extensions similar to halos and are
also detected in merging clusters, but they are usually located in the
cluster outskirts and have an elongated morphology.

\vspace{0.1cm}

\noindent \textbf{Hard X-ray (HXR) observations - } Intracluster
relativistic electrons can give rise to HXR emission through Compton
scattering of CMB photons \cite{1977ApJ...212..608R}. The satellites
Beppo-SAX and RXTE allowed to detect a non-thermal HXR excess in
several clusters hosting diffuse radio sources
(\cite{2007ApJ...654L...9F,2008SSRv..134...71R} and refs.
therein). However, if the radio emission from the non-thermal
component is well established, the presence of an intracluster
non-thermal HXR excess is at present debated. This is firstly due to
the challenging detection of this kind of X-ray emission, complicated
by astrophysical and instrumental backgrounds
\citep[e.g.][]{2004A&A...414L..41R,2004ApJ...602L..73F}. Other
instruments (e.g. INTEGRAL, Swift, Suzaku) are however confirming the
presence of an HXR excess in clusters hosting diffuse radio
sources. The nature of the HXR excess is, in turn, debated.

Several papers agree with the {\it non-thermal origin} of this
emission. In about 10 clusters hosting diffuse radio sources HXR data
are in agreement with the Compton scattering hypothesis
\citep[e.g. the Coma
cluster][]{1999ApJ...511L..21R,1999ApJ...513L..21F,2007arXiv0710.4417E}.
Other observations on the contrary suggest that the non-thermal HXR
excess is due to a population of AGNs
\citep[e.g.][]{2009ApJ...690..367A}: since the hard X-ray instruments
on both Beppo-SAX and RXTE are non imaging, contamination by very hard
point sources inside the cluster cannot be excluded. Two other
non-thermal hypotheses have been suggested from the theoretical point
of view: the HXR excess could be due to non-thermal or supra-thermal
bremsstrahlung \citep[e.g][]{2000ApJ...532L...9B,2008ApJ...675..156W},
or to synchrotron emission from ultrarelativistic electrons
\citep[e.g][]{2004A&A...417..391T}.  Finally, recent observational
results confirm the detection of an HXR excess, but relate it to a
{\it multi-temperature ICM emission}
\citep[e.g.][]{2009ApJ...690..367A,2008A&A...491..363O}.  There are
currently growing observational and numerical evidences pointing
towards a non-uniform distribution of the ICM temperature in merging
clusters, where diffuse radio sources have been detected and shocks
can heat the ICM up to temperatures of several tens of keV
\citep[e.g.][]{2008arXiv0811.0834M}. All these hypotheses need to be
tested both from the theoretical and from the observational point of
view through combined multi-frequency data
\citep[e.g.][]{2008A&A...479...27E}

\subsection{Main physical properties and open questions}\label{sec:pp}

\noindent \textbf{Magnetic fields -} Intracluster magnetic field
intensities can be measured \cite{2004IJMPD..13.1549G}: a) from the
ratio between the synchrotron and Compton HXR fluxes, under the
assumption that they are produced by the same population of
relativistic electrons, b) through Faraday rotation analysis of radio
sources in the background or in the galaxy clusters themselves, c) by
assuming that the energy density of the relativistic plasma within the
diffuse radio source is minimum (equipartition hypothesis), d) from
the X-ray analysis of cold fronts. These measures give quite
discrepant results, with magnetic field values ranging from $\sim$0.1
$\mu$G to $\sim$10 $\mu$G.  These differences can be due to several
factors, such as the various physical assumptions of the different
methods, and the complicated spatial profile and structure of both the
magnetic field and the gas density
\cite{2008SSRv..134...93F}. Variations of the magnetic field structure
and strength with the cluster radius have been recently pointed out by
\citet{2006A&A...460..425G}, stressing the need of more sophisticated
studies of intracluster magnetic fields.

\vspace{0.1cm}

\noindent \textbf{Relativistic particles -} Due to radiative losses,
the typical lifetime of relativistic electrons in the ICM is too short
($\sim$0.1 Gyr) to allow them to propagate with diffusion Alfv\'enic
velocities over the typical Mpc scales of halos and relics. This
requires {\it in situ} injection of relativistic electrons by physical
mechanisms which are usually divided in two main classes: primary and
secondary models. The former predict the (re)acceleration of fossil
radio plasma or directly of thermal electrons of the ICM through
shocks and/or MHD turbulence, which, in turn, can be generated either
by cluster merging (in the case of halos and relics
\citep[e.g.][]{1998A&A...332..395E,2006MNRAS.369.1577C}), or by the
central cooling-flow (in the case of mini halos
\cite{2002A&A...386..456G}). Secondary models predict that non-thermal
electrons are the secondary products of hadronic interactions between
relativistic protons (which have long lifetimes in clusters) and the
ions of the thermal ICM \citep[e.g.][]{1999APh....12..169B}. p-p
collisions take place in all galaxy clusters, both relaxed and
mergers, and are expected to inject also neutral pions in the ICM,
which in turn decay and produce gamma-ray emission.

Most of the theoretical predictions of primary models are in agreement
with current observational results, in particular concerning the
spectral behaviour of radio halos and relics, and the correlation
between the physical properties of these radio sources and of their
host clusters (dynamical state, mass, luminosity, ...; see
\cite{2008SSRv..134...93F} and refs. therein). A certain number of
open questions needs anyway to be addressed, such as: do all (merging)
clusters host diffuse radio sources? Is the low number of known radio
halos and relics related to physical or instrumental effects? What is
the origin of peculiar sources and of the seed electrons for
re-acceleration models? To answer this last question, hybrid models
for electron acceleration (i.e. a combination of primary and secondary
mechanisms) have also been proposed. These works stress the importance
of the next generation of radio (e.g. LOFAR), gamma-ray (e.g. Fermi)
and HXR telescopes (e.g. Simbol-X) to put essential constraints on
current theoretical predictions
\cite{2008MNRAS.385.1211P,2009arXiv0901.1432B}.

\section{Perspectives: LOFAR and Simbol-X}\label{sec:LSX}

Enormous perspectives for the investigation of the non-thermal
intracluster component will be opened by LOFAR (LOw Frequency ARray)
and the Simbol-X satellite \citep[e.g.][]{2008MmSAI..79..176F}. The
latter, with its imaging capabilities up to high energies ($\sim$ 80
keV), will allow to study unambiguously the presence and origin of HXR
emission in clusters \citep[e.g.][]{2008MmSAI..79..170A}. LOFAR is an
array of radiotelescopes that will open the low-frequency spectral
window ($<$300 MHz) to deep and high-resolution observations
\citep[e.g.][]{2003NewAR..47..405R}. This low-frequency domain is
optimal for the detection of high spectral index radio sources, such
as radio halos, mini-halos and relics (see Fig. 7 in
\cite{2008SSRv..134...93F}), and about 1000 halos and relics are
expected to be detected by LOFAR, of which 25\% at z$\gtrsim$0.3
\cite{2003NewAR..47..405R}. Statistical studies of the evolution of
diffuse radio emission in clusters as a function of redshift and
cluster physical properties will thus become possible. A combination
of the LOFAR cluster catalogue with new X-ray and SZ cluster surveys
(e.g. XMM-LSS, Planck, ...) will be suitable for a full understanding
of galaxy cluster physics.

\begin{SCfigure}
   \centering
   \includegraphics[width=2.5in]{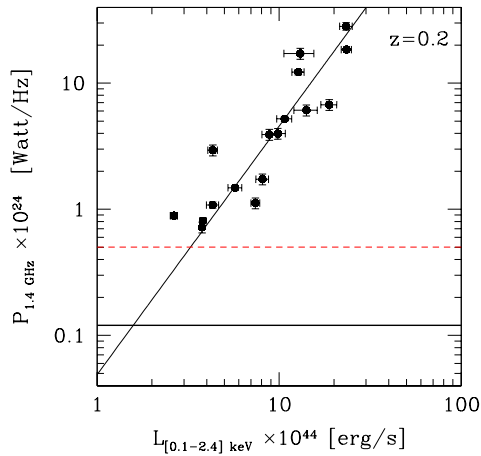} 
   \caption{Radio power of observed halos {\it vs.} X-ray luminosity
     of their host clusters. The two quantities show a strong
     correlation \citep[e.g.][]{2007MNRAS.378.1565C}. The horizontal
     dashed line indicates the approximate limit of current radio
     observations. The limit expected at $z$=0.2 in the LOFAR ``All
     Sky Survey'' for the 10$\sigma$ detection of at least 50\% of the
     flux of radio halos is indicated by the solid line. The observed
     correlation between radio halos' luminosities and sizes, and the
     X-ray luminosity of the cluster have been adopted.}
   \label{fig:fig2}
\end{SCfigure}

It has recently been estimated that the best targets for Simbol-X
studies of the non-thermal intracluster emission are clusters at
$z\lesssim$0.2
\cite{2008MmSAI..79..170A,2008MmSAI..79..182B}. Fig. \ref{fig:fig2}
shows that, in the same redshift range, the LOFAR ``All Sky Survey''
\cite{2007AAS...210.6604R} is expected to detect cluster diffuse radio
sources well below current observational limits. At $z<0.2$, joint
detailed LOFAR/Simbol-X observations will thus allow to study:

\vspace{0.1cm}

\noindent \textbf{- The energy spectrum of relativistic particles:} it
can be complementary constrained by radio and HXR
observations. Actually, if the observed synchrotron radio emission and
Compton HXR emission are produced by the same electron population, they
share the same spectral index $\alpha$ (S$(\nu) \propto
\nu^{-\alpha}$). This spectral index relates to the index $\delta$ of
the power-law electron energy density distribution as $\delta =
2\alpha +1$.

\vspace{0.1cm}

\noindent \textbf{- Intracluster magnetic fields:} their intensity
and, possibly, structure will be measured with a significantly better
statistics through: a) Faraday rotation measures of hundreds of
embedded/background radio galaxies and joint polarisation studies of
diffuse radio sources \cite{GOVONI_EMMEN07} (LOFAR observations); b)
the ratio between the synchrotron and Compton HXR fluxes (see Eq. 13
in \cite{2008SSRv..134...93F}; LOFAR + Simbol/X observations).

\vspace{0.1cm}

\noindent \textbf{- The nature of radio relics:} Simbol-X will be
particularly suited for the study of cluster radio relics detected by
LOFAR. The X-ray emissivity of the thermal ICM depends on the square
density of the gas, and is thus significantly lower in the external
regions of clusters (where relics are located) compared to the central
parts.  That is the reason why the correlation between thermal shocks
and radio relics has been so poorly constrained by observations up to
now \citep[e.g.][]{2006Sci...314..791B}. This decrease of the X-ray
thermal emissivity in the cluster outskirts makes the detection of the
non-thermal component over the thermal emission easier in the relic
regions (for more details, see \cite{2008MmSAI..79..170A}).


\begin{theacknowledgments}
  I am very grateful to the organisers Philippe Ferrando and Paolo
  Giommi and to the Organising Committee for the invitation to this
  interesting and stimulating conference. I warmly thank Christophe
  Benoist and Federica Govoni for their careful reading of the
  manuscript and useful comments.
\end{theacknowledgments}


\bibliographystyle{aipproc}   

\bibliography{c_ferrari}

\IfFileExists{\jobname.bbl}{}
 {\typeout{}
  \typeout{******************************************}
  \typeout{** Please run "bibtex \jobname" to optain}
  \typeout{** the bibliography and then re-run LaTeX}
  \typeout{** twice to fix the references!}
  \typeout{******************************************}
  \typeout{}
 }

\end{document}